# Thermally activated magnetization reversal in bulk $BiFe_{0.5}Mn_{0.5}O_3$

D. Delmonte[1]*, F. Mezzadri[2], C. Pernechele[1], G. Calestani[2-3], G. Spina[4], M. Lantieri[5], M. Solzi[1], R. Cabassi[2], F. Bolzoni[2], A. Migliori[6], C. Ritter[7] and E. Gilioli[2]

[1]*Dipartimento di Fisica, Università di Parma, Parco Area delle Scienze 7/A, 43124 Parma, Italy*

[2]*IMEM-CNR, Parco Area delle Scienze 37/A, 43124 Parma, Italy*

[3]*Dipartimento di Chimica, GIAP Università di Parma, Parco Area delle Scienze 17/A 43124 Parma, Italy*

[4]*Dipartimento di Fisica, Università di Firenze, via Sansone 1, 50019 Sesto Fiorentino (FI), Italy*

[5]*Istituto Sistemi Complessi-CNR, via Madonna del Piano 10, 50019 Sesto Fiorentino (FI), Italy*

[6]*IMM-CNR, via Gobetti 101, 40129 Bologna, Italy*

[7]*Institute Laue-Langevin, Boite Postale 156, F-38042, Grenoble, France*

**\*Corresponding author**

E-mail: davide.delmonte@fis,unipr.it   FAX: +39 0521  269206   TEL: +39 0521 269220

We report on the synthesis and characterization of $BiFe_{0.5}Mn_{0.5}O_3$, a potential type-I multiferroic compound displaying temperature induced magnetization reversal. Bulk samples were obtained by means of solid state reaction carried out under the application of hydrostatic pressure of 6 GPa at 1100°C. The crystal structure is a highly distorted perovskite with no cation order on the B site, where, besides a complex scheme of tilt and rotations of the $TM-O_6$ octahedra, large off-centering of the bismuth ions is detected. Below $T_1 = 420$ K the compound undergoes a first weak ferromagnetic transition related to the ordering of iron rich clusters. At lower temperatures (just below RT) a complex thermally activated mechanism induces at first an enhancement of the magnetization at $T_2 = 288$ K, then a spontaneous reversal giving rise to a negative response. The complementary use of powder neutron diffraction, SQuID magnetometry and Mössbauer spectroscopy allowed the interpretation of the overall magnetic behaviour in terms of an uncompensated competitive coupling between non-equivalent clusters of weakly ferromagnetic interactions characterized by different critical temperatures and resultant magnetizations.

PACS numbers: 75.85.+t, 75.60.Jk, 76.80.+y, 75.30.Et, 75.30.Kz

## I. INTRODUCTION

The members of the $BiFe_{1-x}Mn_xO_3$ solid solution are potentially multiferroic materials, displaying different structural, magnetic and electric properties depending on the x value. Both the end members of the series (x=0,1) are widely studied, due to the presence of multiferroic properties with above room temperature ordering temperatures. $BiFeO_3$ crystallizes in a distorted perovskite structure with R3c rhombohedral symmetry related to the presence of ferroelectric properties. Antiferromagnetism is detected, ascribed to the cycloidal rotation of a spin-canted magnetic structure.[1] On the other hand $BiMnO_3$ is an orbital-order induced ferromagnet with ordering temperature around 100 K.[2,3] The accurate determination of its crystallographic structure is matter of debate, in particular for what concerns the presence of the inversion center; indeed at present the existence of ferroelectricity in $BiMnO_3$ is still under investigation.[4,5] Surprisingly, the members of the Fe/Mn solid solution are poorly studied in bulk form,[6,7,8,9] despite the observation of interesting chemical and physical properties. The most intriguing phenomenon reported so far for the compounds with $0.25<x<0.5$ is the so called spontaneous magnetization reversal (MRV). In several ferrimagnets, as predicted by Neél,[10] MRV occurs caused by the presence of two different temperature dependences of sublattice magnetizations arising from magnetic ions at non-equivalent crystallographic sites;[11,12,13,14,15,16,17,18] while the presence of disorder on the perovskite B site, as for instance in $YFe_{0.5}Cr_{0.5}O_3$,[19] and $(La_{1-x}Bi_x)Fe_{0.5}Cr_{0.5}O_3$,[20] together with Dzyaloshinskii-Moriya (DM) interaction gives rise to MRV due to uncompensated weak ferromagnetism involving clusters of non-equivalent exchange interactions. For what concerns $BiFe_{0.5}Mn_{0.5}O_3$, the current explanations of the phenomenon ascribe the process to a thermally activated competition between DM interaction and single ion anisotropy[6] (hypothesis that is not strengthened by sufficient experimental confirmations), or to an extrinsic process due to the presence of inhomogeneities.[9] In order to clarify the nature of the MRV in the system, we performed Mössbauer measurements on a 48% $^{57}$Fe-enriched $BiFe_{0.5}Mn_{0.5}O_3$ sample, together with accurate structural analyses and magnetization measurements, whose results are presented in this work. The performed study yields experimental support to a mechanism which is at least coexistent (if not alternative) to the previously hypothesized ones. It is shown that the MRV is ascribable to the competitive coupling of iron- and manganese-rich regions characterized by different exchange interactions, resultant magnetizations and critical temperatures.

## II. EXPERIMENTAL

Polycrystalline $BiFe_{0.5}Mn_{0.5}O_3$ was synthesized via a solid state reaction carried out in HP/HT conditions, using a Walker-type multianvil press. The starting powder binary oxides ($Bi_2O_3$, $Fe_2O_3$, $Mn_2O_3$) in stoichiometric amounts were grinded together and encapsulated in gold foils. The best thermodynamic synthesis conditions were determined as 6 GPa of isotropic pressure, 1100°C, and 1.5 hours of reaction time. The sample was quenched to room temperature before pressure was slowly released.

Powder XRD patterns were collected using Cu $K_\alpha$ radiation with a Thermo ARL X'tra powder diffractometer equipped with a Thermo Electron Si(Li) solid state detector to eliminate the incoherent background produced by the florescence of Fe and Mn. Data collections were performed by 0.01-0.02° steps with counting time ranging from 3 to 10 sec.

Single crystal XRD data were collected with Mo $K_\alpha$ radiation on a Bruker AXS Smart diffractometer, equipped with an APEX II CCD area-detector.

Electron diffraction (ED) and high resolution electron microscopy (HREM) were carried out using a Philips TECNAI F20 transmission electron microscope operating at 200 kV. The specimens were prepared by grinding the powders in isopropylic alcohol and evaporating the suspension on a copper grid covered with a holey carbon film.

Powder neutron diffraction data were collected at the D1B and D2B beamlines of the Institut Laue-Langevin in Grenoble. Experiments were performed in the 10-500 K temperature range, with incident beam wavelengths 2.52 and 1.59 Å. All the refinements were carried out using the GSAS package.[21,22]

Magnetic measurements were performed, operating only in standard DC mode, by using a SQuID magnetometer MPMS-XL. The instrument allows the control of both temperature (between 5 and 680 K) and magnetic field intensity (from 0 to 5 T). Supplementary magnetic measurements were performed at higher temperatures (up to 850°C) by using a DSM8 Stationary Pendulum Magnetometer.

Mössbauer measurements were performed by means of a Wiessel spectrometer, calibrated by using a standard metal iron foil, and an Oxford flux cryogenic system with a base temperature of 1.8 K. The source was a 25 mCi $^{57}$Co in Rhodium matrix with Lamb-Mössbauer factor $f_s = 0.63$, as measured by applying the method described in Ref. 23. Twenty-two spectra were collected between 67.8 K and 300 K in the absence of applied field: four of them above the higher critical MRV temperature $T_2 = 288$ K (as explained in detail in the text), five between $T_2$ and $T_3 = 250$ K, which is the second critical MRV temperature, five others from $T_3$ to the compensation temperature at 10 Oe ($T^* = 186$ K) and finally, eight spectra under $T^*$. Since we were interested in evaluating the magnetic components of the spectra, by mixing 48% $^{57}$Fe enriched active material powder with Boron nitride as eccipient we prepared a high-$t_a$ sample, containing 22.08 mg/cm$^2$ of compound.

### III. RESULTS

**Structural analysis**

Preliminary single crystal XRD experiments were initially performed on different samples. In all cases most of the observed reflections could be indexed on the basis of an orthorhombic perovskite superstructure with $a = 5.5728(5) \approx \sqrt{2}a_p$, $b = 11.2065(10) \approx 2\sqrt{2}a_p$, and $c = 15.7430(15)$ Å $\approx 4a_p$, 16 times in volume with respect to the fundamental pseudocubic perovskitic cell with lattice parameter $a_p$. However additional weaker satellites were typically observed in single crystal patterns, suggesting the presence of a larger pseudocubic superstructure with $a \approx b \approx c \approx 4 a_p$. A careful analysis pointed out that the relative intensity of the satellites varies from sample to sample, suggesting a twinning phenomenon at the origin of these weaker reflections. This agrees with the results of powder XRD, reported as Supplemental Material in Fig. 1,[24] where, besides the reflections indicating the presence of the sole extra phase of Bi$_2$CO$_5$ in amounts of few percents, no extra peaks were observed by indexing the pattern with the orthorhombic cell. A definitive confirmation was offered by TEM experiments, indicating the orthorhombic lattice as the true cell of the structure. Selected area electron diffraction (SAED) patterns taken along the fundamental [001] and [010] zone axes of the orthorhombic cell are shown in Fig. 1. The existence of short-scale twinning involving the exchange of the fundamental perovskite axes (quite similar by the metrical point of view) was clearly evidenced by high resolution electron microscopy (HREM) coupled with SAED, the data and their accurate analysis being reported in Fig. 2. Different crystals were analyzed by single crystal XRD in order to identify a "single-domain" sample, but all the examined crystals having suitable dimensions were found to be more or less affected by twinning. In order to solve and refine the structure intensities, data were collected from the crystal showing the lowest twinning contribution. The structure was solved using SIR2004[25] in the *Pnam* space group and refined with SHELX97[26] making use of anisotropic thermal parameters for all atoms. The transition metal sites, TM1 and TM2, were refined by constraining the occupancy of both Mn and Fe to 50%, in agreement with the observed equivalence of their average bond distances. Owing to the presence of residual twinning contributions to the observed intensities, the agreement indices obtained in the refinement are not optimal, but the description of the structure can be considered completely reliable, as confirmed by the Rietveld refinement of the powder diffraction data, produced by using the structural parameters determined by single crystal XRD. Crystal data and refined parameters are reported in Table I (and Table I in Supplemental Material[24]), while selected cation-oxygen bond lengths are gathered in Table II. The cation-oxygen bond lengths evidence large distortions of the coordination around the bismuth atoms, induced by the strong stereochemical activity of the $6s^2$ lone pair. Noteworthy, this effect involves also the iron/manganese coordination octahedra, which appear to be largely distorted and consequently high values of the corresponding quadrupolar splitting are expected (for Fe ions $Q_s$=0.65 mm/s). In spite of this the two TM sites show, as previously pointed out, the same average bond distance, confirming the absence of cation ordering. Analysis of the TM-O distances in terms of charge distribution was performed using the program CHARDIS99,[27] suggesting the exclusive presence of $Fe^{3+}$ and $Mn^{3+}$ within the structure (see Table II). The charge distribution method implemented in CHARDIS99 is a development of the bond-valence approach, allowing a more reliable treatment of atomic charges in solid solutions and structures with distorted coordinations. A representation of the crystal structure projected along the [001] and [100] directions is reported in Fig. 3.

The large superstructure observed in $BiFe_{0.5}Mn_{0.5}O_3$ is quite unusual, as most of the known orthorhombic double perovskites display a crystallographic cell with $a \approx b \approx \sqrt{2}a_p$, $c \approx 2a_p$ related to the tilt of the $BO_6$ octahedra. The large periodicity is ascribed to the distortions produced by the bismuth atoms inducing a complex structure where the $TMO_6$ octahedra are both tilted along the *c* direction, with TM-O-TM bond angles ranging from 157.6° to 146.2°, and rotated in the *ab* plane with a + + - - scheme[28] along c. Noteworthy the displacement scheme of the bismuth ions involves the notable formation of dimers that are consistent with an antiferroelectric structure, as shown in Fig. 4.

Surprisingly, if one considers the dramatically different structures of the two end members of the $BiFe_{1-x}Mn_xO_3$ solid solution, the structure observed here is very close to the one reported in Ref. 7 for $BiFe_{0.75}Mn_{0.25}O_3$, in particular for what concerns the scheme of both octahedral distortions and bismuth ions shifts. Within this framework the observation of similar physical properties in $BiFe_{0.5}Mn_{0.5}O_3$ and $BiFe_{0.75}Mn_{0.25}O_3$ suggests a strong relation structure-properties in this family of compounds.

High resolution neutron diffraction data were collected at 10, 310 and 500 K allowing a thorough refinement of the structural features of the system which points out the absence of symmetry changes in the whole investigated temperature range. Two phases, namely $BiFe_{0.5}Mn_{0.5}O_3$ and $Bi_2CO_5$, in a 20:1 ratio, were necessary to index the patterns. Rietveld plots of the refinements performed at 500 and 10 K are reported in Fig. 5. Thanks to the high resolution data collected at 500 K it was possible to study the sample in its paramagnetic phase confirming the goodness of the previously proposed structural model. In particular, due to the difference in atomic scattering factors of the iron and manganese ions in neutron diffraction, it was possible to exclude the presence of B-site cation ordering. The 10 K data show the raising of purely magnetic peaks with k = (0 0 0) propagation vector related to antiferromagnetic G-type magnetic ordering of the B-site transition metals. The spin arrangement is shown in Fig. 6 and involves collinear atomic moments along the *a* direction, with zero component along *b* and *c*, and the presence of isotropic first neighborhood antiferromagnetic interactions. The observed atomic moment is 5.1 $\mu_B$, slightly higher than the expected one (4.5 $\mu_B$) for an averaged structure containing high-spin $Fe^{3+}$ and $Mn^{3+}$ ions, indicating that the magnetic structure at 10 K is long-range ordered and clearly shows the absence of inhomogeneities involving the magnetic structure and sizeable spin fluctuations.

The high-flux data collected at D1B between 10 and 500 K allowed the accurate study of the magnetic structure thermal evolution. No spin reorientations as well as changes in the propagation vector were observed in the whole examined temperature range. Fig. 7 reports the refined atomic moment as a function of temperature, showing a Brillouin-like behavior up to about 290 K and suggesting the loss of long range magnetic correlation to be located at this temperature. However a weak and broad signal, observed in the 2θ regions where the magnetic reflections are detected, persists up to about 400 K, indicating the presence of small clusters of magnetically ordered TM ions. A similar behavior, as well as G-type spin structure, was previously observed in $YFe_{1-x}Mn_xO_3$, being ascribed to short range magnetic ordering.[29]

The thermal evolution of the lattice parameters, reported in Fig. 8, shows an appreciable anisotropic behavior. Differently from *a*, that shows a linear behavior at high temperatures, both the *b* and *c* parameters display anomalies at 288 K, i.e. the long-range magnetic ordering temperature; the anomalies consist in opposite sign slope variations, slight for *b* but noticeable for *c*, suggesting the presence of spin-lattice coupling, which could be indicative of magnetoelectric effects.

**Magnetic characterization**

Field cooling (FC) magnetometry was performed between 5 and 680 K with applied field H = 100 Oe on an as-grown pellet. The measurements, recorded both on cooling and warming, are reported in Fig. 9. The sample shows a primary paramagnetic-to-weak ferromagnetic transition at $T_1$ = 420 followed at lower temperature by a complex mechanism composed by two interconnected phenomena taking place at $T_2$ = 288 K and $T_3$ = 250 K, finally leading to MRV. High temperature measurements aimed at the study of the paramagnetic region of the compound were performed with applied field H = 10 kOe, due to the weak magnetic signal of the sample and are reported in the inset of Fig. 9. The fitting of the linear region of the inverse susceptibility curve (above $T_1$) allowed to determine the Curie-Weiss temperature θ = -400 K, revealing a global antiferromagnetic nature of the interactions, and a number of Bohr magnetons per formula unit corresponding to 5.4 $\mu_B$. This value is consistent if compared to the expected value 5.2 $\mu_B$ for high-spin $Fe^{3+}$ (5.9 $\mu_B$) and $Mn^{3+}$ (4.9 $\mu_B$). The observed weak ferromagnetic signal between $T_1$ and $T_2$ can be ascribed to a second-order mechanism allowed by the low symmetry of the system and producing spin canting, as for example single ion anisotropy or antisymmetric DM interaction. Just below room temperature the low-field susceptibility undergoes a two-step transition (see Fig. 9): at $T_2$ antiferromagnetic long range order takes place and below $T_3$ a complex thermal mechanism leads the system at first to a compensation ($\chi = 0$) and then to a remarkable negative response. On the contrary, high-field measurements do not show the same mechanism of compensation and at $T_3$ the transition disappears, suppressed by the applied magnetic field, as shown as Supplemental Material in Fig. 2.[24] This suggests the interactions leading to MRV to be very weak, in agreement with the hypothesis of weak ferromagnetism induced by a second-order mechanism. In a recent paper,[9] the reversal of magnetization in $BiFe_{1-x}Mn_xO_3$ (x = 0.3, 0.4) is considered an extrinsic process due to the presence of inhomogeneities related to the observation of Exchange Bias in the hysteresis loop and the absence of MRV, if the system is Field Cooled below $T_N$ ($T_2$ in this case). Wide magnetic characterizations we performed are in contrast with these results, as resumed in the Supplemental Material in Figs. 3-4.[24]

M(H) measurements performed at 300 K, 280 K, 240 K, 100 K and 5 K are reported in Figs. 10a-b. The hysteresis loops were measured starting from 5 T up to -5 T after a field cooling in low magnetic field. All the collected data show that the high field magnetization trend is typical of an AFM system; however in the low field regime two different behaviors, taking place below and above $T_3$ respectively, can be discriminated. At higher temperatures the hysteresis is generated by the weak ferromagnetic component and gives rise to a small remanent magnetization (0.012 emu/g), that could be explained by the presence of mesoscopic structures (clusters) of spin-canted domains. Below $T_3$, the coercive field and spontaneous magnetization increase and the hysteresis loop starts opening out, showing a complex shape, which is probably due to the coexistence of two different weak ferromagnetic contributes (see the blue circle of Fig. 10b), whose competition is possibly responsible for MRV. At 5 K the presence of two symmetrical kinks (Fig. 10a) suggests for the two components a "soft" and a "hard" character respectively.

Another remarkable characteristic of the M(H) curves is the increase, below $T_3$, of the high field susceptibility by decreasing the temperature. Even if quite unusual for magnetically ordered compounds, this behavior could be expected for systems where MRV originates from the competition of two independent magnetic components with different exchange constants, the weaker dominating at lower temperature. The obtained value of remanent magnetization at 5 K is 0.062 $\mu_B$/F.U., consistent with a spin-canted system, while the coercive field is 3000 Oe.

**Mössbauer characterization**

Mössbauer spectra collected at different temperatures in the range 293-68 K are reported in Figs. 11a-b. At high T values, spectra are marked by a saturated, winged doublet, which arises from an unsaturated magnetic structure, extending from -8 to +8 mm/s, corresponding to hyperfine fields ≈ 45 T. By decreasing the temperature, the progressive disappearance of the wells between the magnetic lines indicates the growing of medium field components (≈ 25 T), rising from the simultaneous fall of the doublet intensity: around T = $T_3$ the external lines are flat and the depth of the doublet lines is comparable to the ones of the magnetic lines. A further decrease of T affects the lineshape making the external lines to become well defined again, suggesting a falloff in the medium field components (see Fig. 11b). Therefore, the evolution of the Mössbauer cross-section is described by means of a doublet (sub-5) and four magnetic components, characterized by Gaussian field distributions, which lead to sextets of Voigt profiles.[30,31] One of them (sub-4) is used to fit the wings of the doublet and it should be considered as that part of the superparamagnetic-like component which takes into account inhomogeneous and/or homogeneous effects. The three remaining magnetic structures describe the magnetically ordered part of the cross-sections; namely, sub-3 describes the medium field component and a strong hyperfine mean field variation starting around 230 K, and the last two (sub-2 and sub-1) are necessary to outline the high field components, since the low temperature spectra show asymmetric external lines. All of these subspectra are also able to take into account inhomogeneous broadenings of the electric parameters, i.e. quadrupolar splitting and isomer shift, by means of an additional Gaussian distribution, whose line width returns to values lower than $2\Gamma_N$=0.2 mm/s. The electric parameters of the magnetic sextet used to fit the wings of the doublet were obviously constrained to the same values of sub-5 component. The fitting procedure was performed using the transmission integral method, where the reduced $f_s$ factor was estimated throughout additional PHA measurements.[23] The obtained RT values for the chemical shift of all the subspectra are in 0.25-0.35 mm/s range; as far as the quadrupolar splitting is concerned, the doublet and the associated magnetic component have $Q_s$ = 0.65 mm/s while all the other magnetic subspectra have $Q_s$ < 0.30 mm/s. Moreover, for what concerns subspectrum 4, the angle between the hyperfine field direction and the principal axis of the electric field gradient tensor results to be ≈ 0.8 rad. The obtained parameters for T = 68, 120, 200, 240, 285 and 293 K are reported in Table IV, and the data confirm that the iron oxidation state is 3+. All the fields but sub-4 grow by decreasing T. Sub-1 can be associated to the high temperature order of the system preexistent at T > $T_2$, while sub-2 interprets the low temperature magnetic transition since it shows a discontinuity in the thickness and in the hyperfine mean field at $T_2$, as shown in Fig. 12a. Moreover around $T_3$, where the mechanism of reversal is thermally activated, the plot shows a substantial quenching of the sub-3 thickness and a maximization of the sub-2 one. The temperature dependence of the Gaussian broadenings for all the magnetic subspectra is illustrated in Fig. 12b. Subspectrum 4 is characterized by the highest relative variation and a Gaussian broadening close to the mean field value at 230 K. The Gaussian broadenings of the other components are of the order of few tesla and they decrease smoothly with T, making the spectral lines sharper.

The magnetic fields distributions are derived from the parameters reported in Figs. 12a-b and they are illustrated in Fig. 13 for selected temperatures distinguished by strong and weak superparamagnetic-like components, respectively. In correspondence with the high fields region the high-T distributions are characterized by a peak emerging from a flat trend. By lowering T the distributions become double peaked and finally they tend to collapse into a single peak. The main feature in the line shape thermal evolution concerns the thickness of the superparamagnetic-like doublet with its wings (sub-5 and sub-4), which decreases exponentially with $T$ from 50% of the total thickness to a practically negligible percentage under 200 K. A second element to be noted in the Mössbauer spectra thermal evolution regards the hyperfine field mean values of the sub-1 component, the relative thickness of which is almost 20% of the total thickness over 200 K. Although its values are a few tesla smaller, the thermal trend of $B_{hyp}^{(1)}$ follows the one of BiFeO$_3$[32,33] suggesting that the respective transition temperatures are likely to be close to each other.

## IV. DISCUSSION

The synergic use of crystallographic, magnetic and Mössbauer characterizations allowed the comprehension of the complex magnetic behavior in BiFe$_{0.5}$Mn$_{0.5}$O$_3$. The structural data give the fundamental starting point: the iron and manganese ions result to be disordered at the perovskites B site, and the analysis of the bond lengths, of the paramagnetic portion of the $1/\chi$ curve and the iron quadrupolar splitting and isomer shift suggest both the transition metals to be in 3+ oxidation state. The magnetic structure is long-range G-type AFM, therefore all the exchange interactions (Fe-Fe, Fe-Mn, Mn-Mn) are antiferromagnetic, as confirmed by the largely negative Curie-Weiss temperature observed. As a consequence the weakly ferromagnetic moment detected (0.062 $\mu_B$/F.U. at 5 K) should be ascribed to a second order interaction giving rise to spin canting, as for example DM interaction or single ion anisotropy. This is not surprising as several ABO$_3$ perovskite structures with trivalent A ions and B = iron, chromium, manganese and their solid solutions often show DM-induced weak ferromagnetism, as for example the parent compound BiFeO$_3$, where spiral magnetic ordering is observed.[34] Within this framework, the lack of weak ferromagnetic components in the collected neutron diffraction patterns is ascribed to the sensitivity of the technique, not allowing the detection of signals lower than 0.5 $\mu_B$ per magnetic ion. The key for the interpretation of the magnetic behavior is given by the Mössbauer data, showing at 300K the presence of an ordered component (about 20% of the iron atoms) characterized by high hyperfine field values, together with partially ordered states and a superparamagnetic-like doublet; the latter shows an exponential decrease of the thickness as the temperature is decreased, disappearing at 200 K. Since BiFeO$_3$ and BiMnO$_3$, display very different magnetic ordering temperatures (643 and 100 K respectively), it is likely to consider that also in a disordered solid solution the iron- and manganese-rich regions may order at different temperatures, the first of them giving rise to the Mössbauer sub-1 component (Fig. 12a). This interpretation also agrees with the broad and weak signal observed in neutron diffraction above the ordering temperature $T_2$, suggesting the presence of short-range magnetically ordered diffraction domains. Since above $T_1$=420 K the compound is paramagnetic, the 420-288K magnetic behavior is ascribed to a confined G-type antiferromagnetic spin arrangement affected by spin-canted weak ferromagnetism, involving solely the Fe-rich clusters. At $T_2$ the long range magnetic ordering involving both Fe and Mn takes place, driven at first by the preordered iron rich clusters (sub-2 of Fig. 12a), and is observed in the magnetization measurements as an increase of the pre-existent weak ferromagnetic component. However, by decreasing the temperature, the Fe-Mn interactions, which are dominant from a statistical point of view in the equimolar solid solution, become competitive with the pre-existent component and, being evidently connected to a spin canting in the opposite direction, give rise to compensation at $T_3$ and then to spontaneous reversal of the magnetization. Noteworthy is the fact that the observation of the reversal process is always related to the initial orientation at RT of the small resultant moment of the iron-rich clusters. Indeed in the FCC measurements the application of a field forces the weak ferromagnetic component of the ordered clusters to align to positive values; by lowering the temperature the remaining iron ions start to condense into ordered states, assisted by the manganese ones, finally developing negative magnetization.

## V. CONCLUSIONS

The use of structural characterization techniques, magnetization measurements and Mössbauer spectroscopy allowed the meticulous study of the fundamental properties of $BiFe_{0.5}Mn_{0.5}O_3$ in bulk form, making possible the interpretation of the unusual spontaneous (MRV) phenomenon observed in this potentially magnetoelectric compound. The structural characterizations point out that $BiFe_{0.5}Mn_{0.5}O_3$ crystallizes in an orthorhombic perovskite superstructure with a ≈ √2 $a_p$, b ≈ 2√2 $a_p$, c ≈ 4 $a_p$. No cation ordering involving $Fe^{3+}$ and $Mn^{3+}$ is observed and the large distortion of the perovskite lattice is due to a complex scheme of tilt and rotations of the $TM-O_6$ octahedra together with an extended path of displacements of the bismuth ions yielding an antiferroelectric arrangement. Taking into account the disordered nature of the $BiFe_{0.5}Mn_{0.5}O_3$ solid solution, a mechanism able to describe the MRV process has been developed on the basis of Mössbauer, magnetic and neutron diffraction measurements. Basing on the completely different ordering temperatures of the solid solution end-members, the mechanism implies, by decreasing the temperature, a composition dependent progressive ordering of clusters (revealed as antiferromagnetic G-type by neutron diffraction) that starts with the iron-rich ones and then extends on the whole solid. The presence of a weak ferromagnetic component is ascribed to spin-canting, whereas the MRV to the fact that the Fe-Mn interactions, statistically dominant but taking place at lower temperatures, produces a ferromagnetic component which order antiparallel to the preexistent one. It is quite interesting to note that the structural characterization pointed out anomalies in the thermal dependence of the lattice parameters occurring at the magnetic long-range ordering temperature $T_2$, indicating a spin-lattice coupling that could be indicative of magnetoelectric coupling.

## V. ACKNOWLEDGMENTS

Cariparma Credit Agricole is thanked for financial support. Ente Cassa di Risparmio di Firenze is thanked for its financial support (Contract No. 2010.0419). The authors acknowledge the Institute Laue-Langevin (Grenoble, France) for providing technical and financial support.

**Table Captions**

TABLE I. Crystal data and refined parameters.

TABLE II. List of relevant bond distances (Å).

TABLE III. Mössbauer parameter values at selected temperatures. $\delta(i)$: center shift of subspectra with respect to CoRh source, $\delta$: center shift of the spectrum, $Q_s(i)$: quadrupolar splitting of subspectra, $B_{hyp}(i)$: mean value of subspectra hyperfine magnetic fields , $\theta_B(i)$: polar angle of $B_{hyp}(i)$ with respect to the EFG frame, $t_a(i)$: inhomogeneous Gaussian distribution of electric parameters, $\sigma_B$: standard deviation value for the Gaussian distribution of hyperfine magnetic field Mössbauer thickness of subspectra, $t_a$: total Mössbauer thickness, $\sigma_\delta$: standard deviation value for the centered on $B_{hyp}$, The standard deviation for the free parameters are reported in parentheses.

TABLE I.

| | | | | | |
|---|---|---|---|---|---|
| Space group | Pnam | | Reflections collected | | 16500 |
| Unit cell dimensions | a = 5.5728(5) Å | | Data/restraints/parameters | | 1356/0 /97 |
| | b = 11.2065(10) Å | | Goodness-of-fit on $F^2$ | | 1.249 |
| | c = 15.7430(15) Å | | R indices [I>4σ(I)] | | $R_1$ = 0.0561, $wR_2$ = 0.1291 |
| Volume | 983.175 (3) Å$^3$ | | R indices (all data) | | $R_1$ = 0.0641, $wR_2$ = 0.1249 |

| | X | Y | z | s.o.f | $U_{eq}$(Å$^2$) |
|---|---|---|---|---|---|
| Bi1 | 0.27114(17) | 0.13754(8) | ¾ | 1 | 0.0105(2) |
| Bi2 | 0.71523(18) | -0.12525(8) | ¾ | 1 | 0.0107(2) |
| Bi3 | 0.72410(16) | -0.12533(6) | 0.51022(5) | 1 | 0.0189(2) |
| TM1* | 0.7429(4) | 0.12118(19) | 0.62580(14) | 1 | 0.0082(5) |
| TM2* | 0.74827(4) | -0.37683(19) | 0.62344(14) | 1 | 0.0055(5) |
| O1 | 0.680(3) | 0.0769(18) | ¾ | 1 | 0.0144(4) |
| O2 | 0.578(2) | -0.2236(11) | 0.6307(8) | 1 | 0.013(3) |
| O3 | 0.723(4) | 0.1638(16) | 0.5057(9) | 1 | 0.035(4) |
| O4 | -0.032(2) | 0.2473(11) | 0.8433(8) | 1 | 0.010(2) |
| O5 | 0.484(3) | 0.0052(14) | 0.5977(9) | 1 | 0.021(3) |
| O6 | 0.971(3) | -0.0198(15) | 0.6134(10) | 1 | 0.025(4) |
| O7 | 0.815(4) | -0.3877(19) | ¾ | 1 | 0.020(4) |

*site occupancy factors of both Mn and Fe (TM1 and TM2) fixed to 0.50

TABLE II.

| | | | | | | | | | | | |
|---|---|---|---|---|---|---|---|---|---|---|---|
| TM1-O4 | 1.952(13) | TM2-O6 | 1.939(16) | Bi1-O4 | 2.242(13) | Bi2-O7 | 2.240(3) | Bi3-O2 | 2.339(13) | | |
| TM1-O3 | 1.955(15) | TM2-O2 | 1.966(13) | Bi1-O4 | 2.243(13) | Bi2-O1 | 2.27(3) | Bi3-O3 | 2.395(18) | | |
| TM1-O5 | 1.993(16) | TM2-O5 | 1.988(15) | Bi1-O1 | 2.378(18) | Bi2-O2 | 2.308(13) | Bi3-O5 | 2.414(16) | | |
| TM1-O6 | 2.038(16) | TM2-O7 | 2.030(15) | Bi1-O4 | 2.554(13) | Bi2-O2 | 2.308(13) | Bi3-O6 | 2.435(17) | | |
| TM1-O1 | 2.048(14) | TM2-O3 | 2.089(14) | Bi1-O4 | 2.554(13) | Bi2-O6 | 2.838(17) | Bi3-O5 | 2.458(16) | | |
| TM1-O4 | 2.179(14) | TM2-O2 | 2.156(13) | Bi1-O7 | 2.81(3) | Bi2-O6 | 2.838(17) | Bi3-O3 | 2.54(3) | | |
| | | | | Bi1-O5 | 3.059(15) | Bi2-O7 | 2.99(3) | Bi3-O6 | 3.053(17) | | |
| | | | | Bi1-O5 | 3.059(15) | Bi2-O5 | 3.090(16) | Bi3-O3 | 3.12(3) | | |
| | | | | Bi1-O1 | 3.24(3) | Bi2-O5 | 3.090(16) | Bi3-O4 | 3.176(13) | | |
| | | | | Bi1-O6 | 3.245(17) | Bi2-O2 | 3.238(13) | Bi3-O2 | 3.218(13) | | |
| | | | | Bi1-O1 | 3.245(17) | Bi2-O2 | 3.238(13) | Bi3-O3 | 3.241(18) | | |
| | | | | Bi1-O1 | 3.364(17) | Bi2-O7 | 3.350(3) | Bi3-O4 | 3.423(13) | | |
| Avg. | 2.028(14) | Avg. | 2.028(14) | Avg. | 2.833(18) | Avg. | 2.817(19) | Avg. | 2.818(17) | | |
| Q(ij)[a] | 3.11 | Q(ij) | 3.03 | Q(ij) | 2.88 | Q(ij) | 2.73 | Q(ij) | 3.07 | | |

[a] Q(ij)s are the cationic charges as computed by CHARDIS99[27].

TABLE III.

| $T$ (K) | $\chi^2$ | $\delta$ (mm/s) | $t_a$ | $i$ | $\delta(i)$ (mm/s) | $Q_s(i)$ (mm/s) | $B_{\text{hyp}}(i)$ (T) | $\theta_B(i)$ | $t_a(i)$ | $\sigma_\delta(i)$ (mm/s) | $\sigma_B(i)$ (mm/s) |
|---|---|---|---|---|---|---|---|---|---|---|---|
| 68  | 1189 | 0.400 | 12.7 | 1 | 0.387(3)  | 0.150(4)  | 50.56(4) |         | 4.0(8)   | 0.18(1)  | 1.2(1)  |
|     |      |       |      | 2 | 0.403(3)  | 0.166(5)  | 48.0(4)  |         | 5.(1)    | 0.19(1)  | 2.3(2)  |
|     |      |       |      | 3 | 0.417(9)  | 0.20(2)   | 43.0(8)  |         | 2.9(4)   | 0.19(1)  | 5.3(3)  |
|     |      |       |      | 4 | 0.39(3)   | 0.65      | 16.6(5)  | 0.84(8) | 0.51(3)  | 0.15(2)  | 4.8(6)  |
|     |      |       |      | 5 | 0.39      | 0.65      |          |         | 0.06(1)  | 0.15     |         |
| 120 | 1079 | 0.384 | 12.3 | 1 | 0.363(4)  | 0.113(9)  | 49.20(7) |         | 2.5(4)   | 0.15(1)  | 1.55(8) |
|     |      |       |      | 2 | 0.389(3)  | 0.185(6)  | 46.5(2)  |         | 5.5(9)   | 0.22(1)  | 2.8(2)  |
|     |      |       |      | 3 | 0.398(6)  | 0.19(1)   | 41.(1)   |         | 3.5(7)   | 0.18(1)  | 5.1(5)  |
|     |      |       |      | 4 | 0.35(3)   | 0.65      | 17.9(8)  | 0.80(9) | 0.67(7)  | 0.15(3)  | 8.(1)   |
|     |      |       |      | 5 | 0.35      | 0.65      |          |         | 0.08(3)  | 0.15     |         |
| 200 | 1055 | 0.350 | 11.5 | 1 | 0.303(9)  | 0.02(2)   | 46.7(1)  |         | 1.1(2)   | 0.14(2)  | 2.0(1)  |
|     |      |       |      | 2 | 0.353(3)  | 0.171(8)  | 41.8(7)  |         | 5.3(8)   | 0.20(1)  | 4.0(2)  |
|     |      |       |      | 3 | 0.347(7)  | 0.200(2)  | 34.(1)   |         | 3.(1)    | 0.22(1)  | 4.1(8)  |
|     |      |       |      | 4 | 0.366(1)  | 0.65      | 23.(2)   | 0.84(3) | 2.2(5)   | 0.18(1)  | 8.(1)   |
|     |      |       |      | 5 | 0.366     | 0.65      |          |         | 0.25(2)  | 0.18     |         |
| 240 | 1009 | 0.328 | 11.0 | 1 | 0.297(7)  | 0.05(2)   | 44.26(8) |         | 1.5(1)   | 0.18(2)  | 2.6(2)  |
|     |      |       |      | 2 | 0.337(6)  | 0.21(1)   | 34.8(5)  |         | 5.6(8)   | 0.21(1)  | 6.5(4)  |
|     |      |       |      | 3 | 0.38(5)   | 0.24(9)   | 23.6(7)  |         | 0.3(2)   | 0.1(1)   | 3.(1)   |
|     |      |       |      | 4 | 0.325(6)  | 0.64(1)   | 17.(4)   | 0.79(4) | 3.0(9)   | 0.13(1)  | 10.(2)  |
|     |      |       |      | 5 | 0.325     | 0.64      |          |         | 0.53(7)  | 0.13     |         |
| 260 | 989  | 0.309 | 10.6 | 1 | 0.285(7)  | 0.04(2)   | 42.9(1)  |         | 1.9(3)   | 0.20(2)  | 3.1(2)  |
|     |      |       |      | 2 | 0.318(8)  | 0.17(1)   | 30.7(6)  |         | 4.7(8)   | 0.20(1)  | 7.(1)   |
|     |      |       |      | 3 | 0.296(4)  | 0.18(6)   | 22.6(7)  |         | 0.4(4)   | 0.3(1)   | 2.(2)   |
|     |      |       |      | 4 | 0.312(4)  | 0.658(6)  | 13.(1)   | 0.81(4) | 2.2(5)   | 0.16(1)  | 6.(1)   |
|     |      |       |      | 5 | 0.312     | 0.658     |          |         | 1.4(1)   | 0.16     |         |
| 285 | 1004 | 0.297 | 9.9  | 1 | 0.289(7)  | 0.08(1)   | 40.3(1)  |         | 2.57(7)  | 0.16(1)  | 4.0(1)  |
|     |      |       |      | 2 | 0.31(2)   | 0.18(4)   | 31.0(4)  |         | 0.5(3)   | 0.16(1)  | 2.4(6)  |
|     |      |       |      | 3 | 0.32(1)   | 0.13(3)   | 22.6(7)  |         | 2.1(7)   | 0.04(4)  | 5.(1)   |
|     |      |       |      | 4 | 0.291(2)  | 0.632(4)  | 9.(1)    | 0.81(4) | 2.1(4)   | 0.13(1)  | 6.0(8)  |
|     |      |       |      | 5 | 0.291     | 0.632     |          |         | 2.64(5)  | 0.13     |         |
| 293 | 1221 | 0.296 | 9.9  | 1 | 0.281(4)  | 0.089(7)  | 40.4(1)  |         | 2.2(1)   | 0.15(1)  | 3.78(7) |
|     |      |       |      | 2 | 0.38(3)   | 0.24(4)   | 31.5(3)  |         | 0.3(2)   | 0.11(6)  | 2.9(7)  |
|     |      |       |      | 3 | 0.323(9)  | 0.18(2)   | 22.3(6)  |         | 2.7(4)   | 0.       | 8.(1)   |
|     |      |       |      | 4 | 0.28(1)   | 0.63(9)   | 7.1(5)   | 0.82(3) | 1.3(2)   | 0.10(7)  | 4.2(4)  |
|     |      |       |      | 5 | 0.283(1)  | 0.632(1)  |          |         | 3.37(9)  | 0.109(2) |         |

**Figure Captions**

FIG. 1. SAED patterns taken on single domain regions of BiFe$_{0.5}$Mn$_{0.5}$O$_3$ samples.

FIG. 2. HREM image taken in a [111] projection of the fundamental perovskite structure showing the presence of twinning domains involving a 60° rotation of the orthorhombic superstructure lattice around the zone axis. The corresponding experimental SAED pattern is shown in inset a), compared with the FFT obtained on the whole as well as on the upper and lower regions of the image reported in the insets b), c) and d), respectively; c) and d) patterns are indexed on the basis of the orthorhombic superstructure. The "composite" experimental SAED pattern can be explained on the basis of twinning domains evidenced by the corresponding HREM image and its fast Fourier transform (FFT).

FIG. 3. (Color Online) BiFe$_{0.5}$Mn$_{0.5}$O$_3$ crystal structure projected on the *ab* (top) and *bc* (bottom) planes made using the program VESTA.[35] The vertex-sharing light green octahedra are linked by the red oxygen atoms, while the purple spheres represent the bismuth ions.

FIG. 4. (Color Online) Relative shifts of the bismuth ions in the crystal structure of BiFe$_{0.5}$Mn$_{0.5}$O$_3$ projected on the *bc* (top) and *ac* (bottom) plane.

FIG. 5. (Color Online) Rietveld refinements of the neutron diffraction patterns collected at 500 and 10 K (top and bottom respectively).

FIG. 6. (Color Online) Magnetic structure of BiFe$_{0.5}$Mn$_{0.5}$O$_3$: in orange are represented the transition metal ions, in purple the bismuth ones, and in red oxygen. The blue and green arrows indicate the antiferromagnetic atomic moments.

FIG. 7. Refined atomic moment in the 10-500 K range. The inset shows the 280-450 K region, where short range magnetic interactions are detected.

FIG. 8. (Color Online) Lattice parameters variation as a function of temperature. Solid lines are guides to the reader's eye. The dashed vertical line indicates the long-range magnetic ordering temperature.

FIG. 9. (Color Online) BiFe$_{0.5}$Mn$_{0.5}$O$_3$ mass susceptibility measured in warming (FCW) and cooling (FCC) with applied field H = 100 Oe. In the inset the inverse of the magnetic susceptibility collected at 10000 Oe.

FIG. 10. (Color Online) (a) Hysteresis loops measured at different temperatures from RT to 5 K across the reversal process thermal thresold. At 5 K a complex kinked shape is detected. (b) Hysteresis loops at 100, 240, 280 e 300 K. were highlighted. At low temperature it is clearly observed the presence of a sensible stricture in the central part of the curve possibly indicating the superimposition of two different magnetic contributions.

FIG. 11. (Color Online) Mössbauer spectra collected (a) above and (b) below T$_3$.

FIG. 12. (Color Online) (a) Thermal trend of magnetic hyperfine mean fields for the five subspectra (the areas of the dots are proportional to the corresponding Mössbauer thicknesses); (b): Gaussian width of the magnetic field distributions vs T. The legend refers to both the graphics.

FIG. 13. (Color Online) Hyperfine magnetic field density at selected temperatures in the low (a) and high (b) $T$ ranges, corresponding to strong and weak superparamagnetic-like components. The different scales for the vertical axis reflect the collapse of the superparamagnetic-like structure.

Figure 1

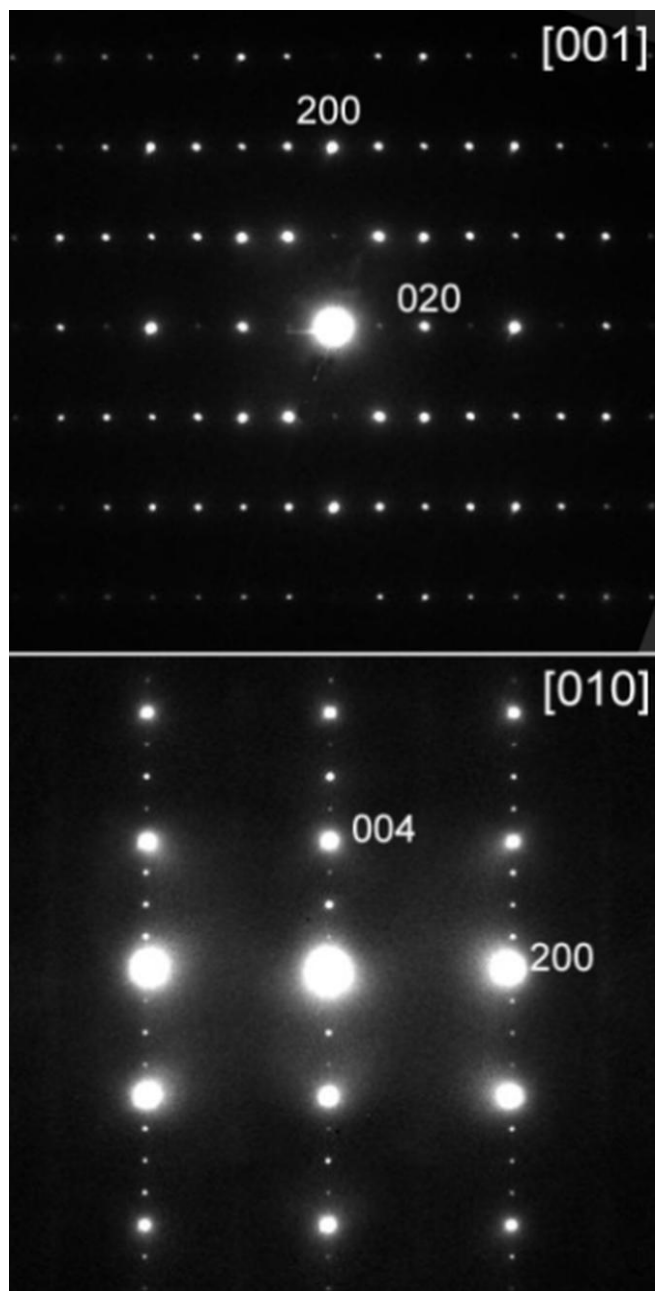

Figure 2

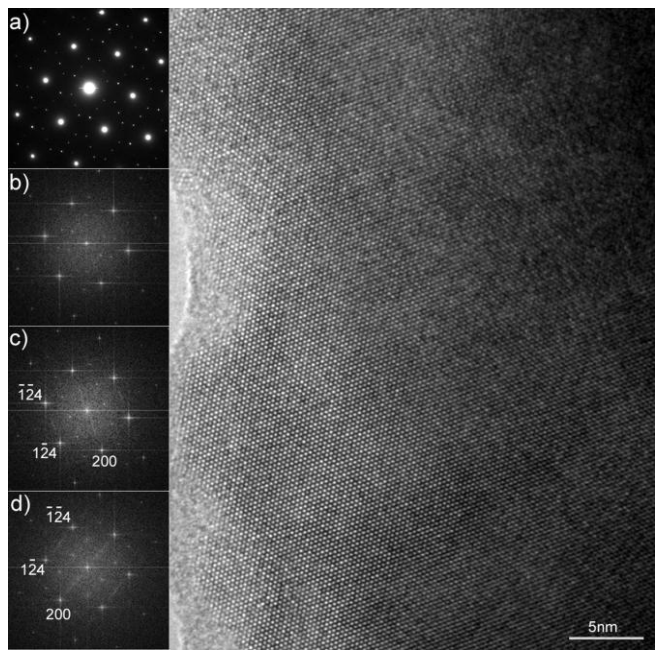

Figure 3

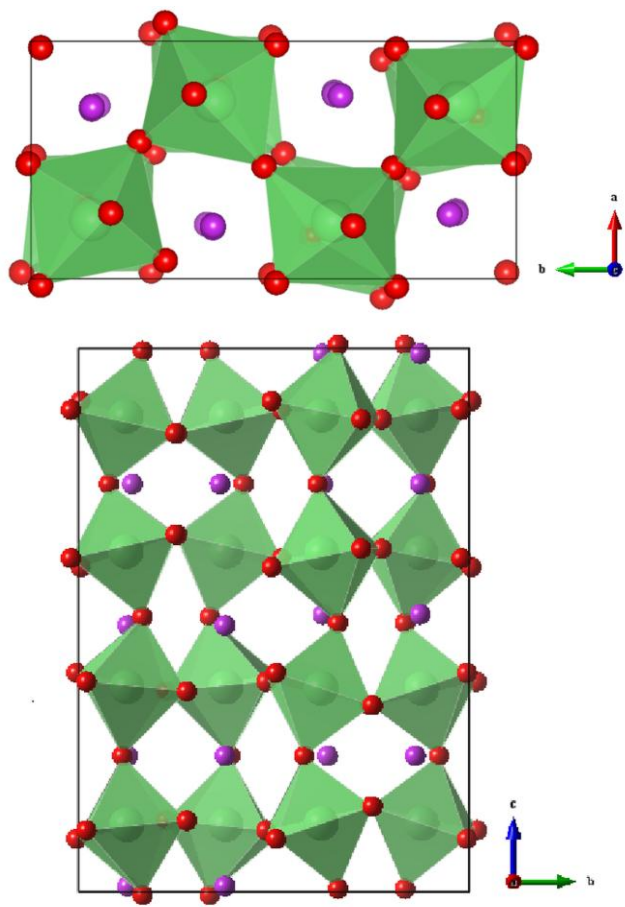

Figure 4

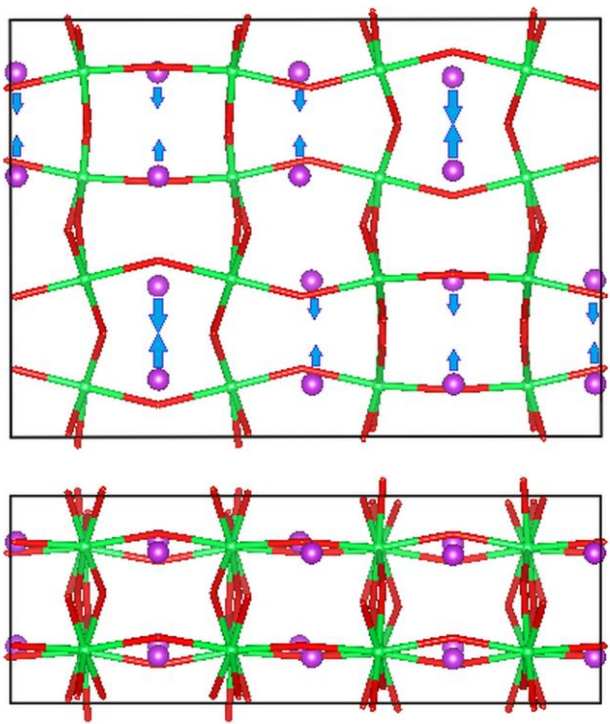

Figure 5

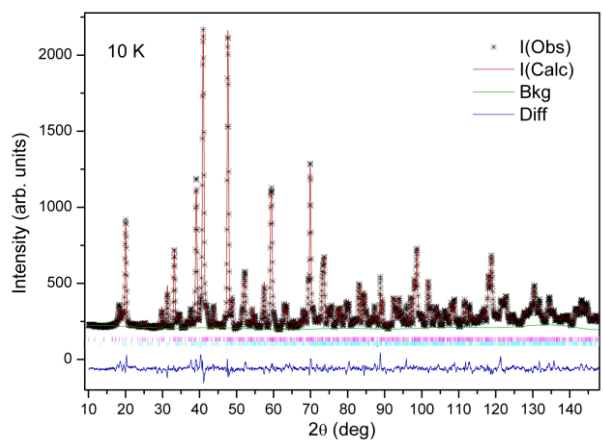

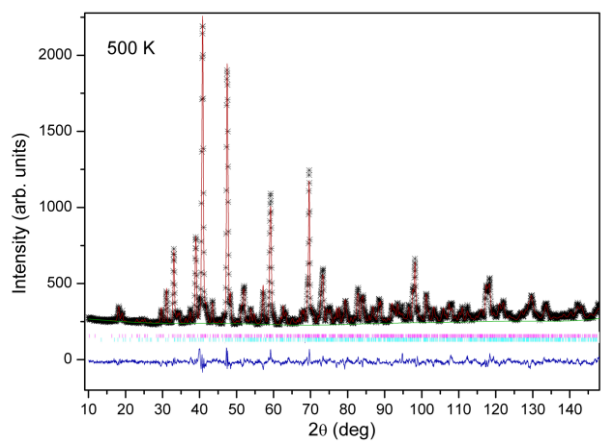

Figure 6

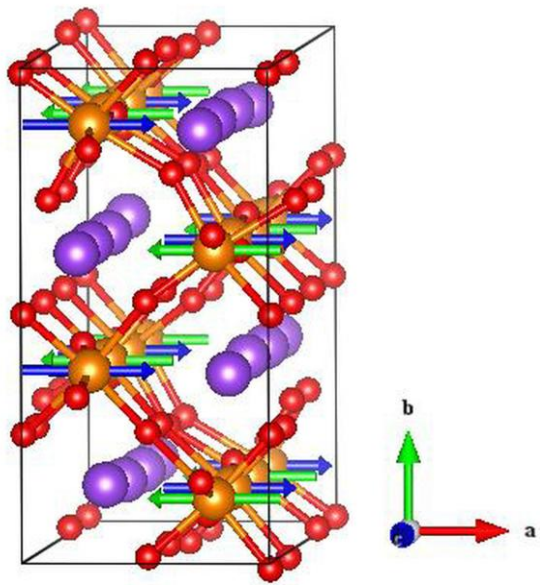

Figure 7

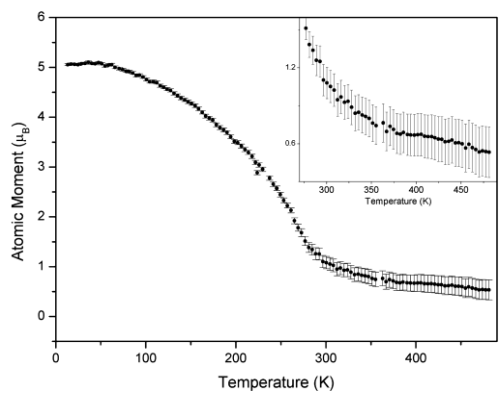

Figure 8

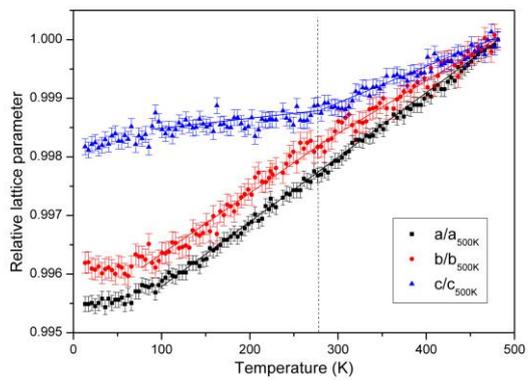

Figure 9

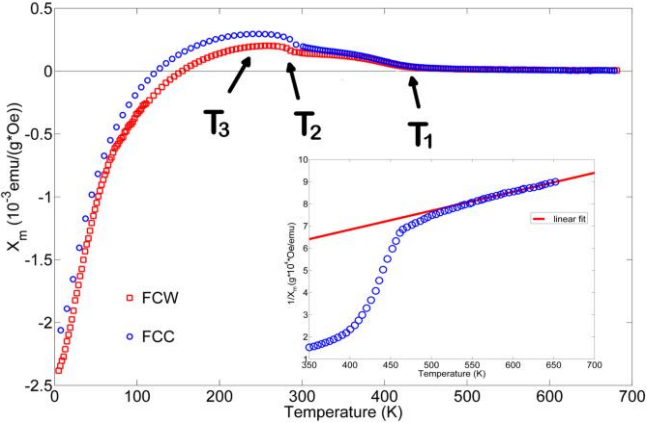

Figure 10

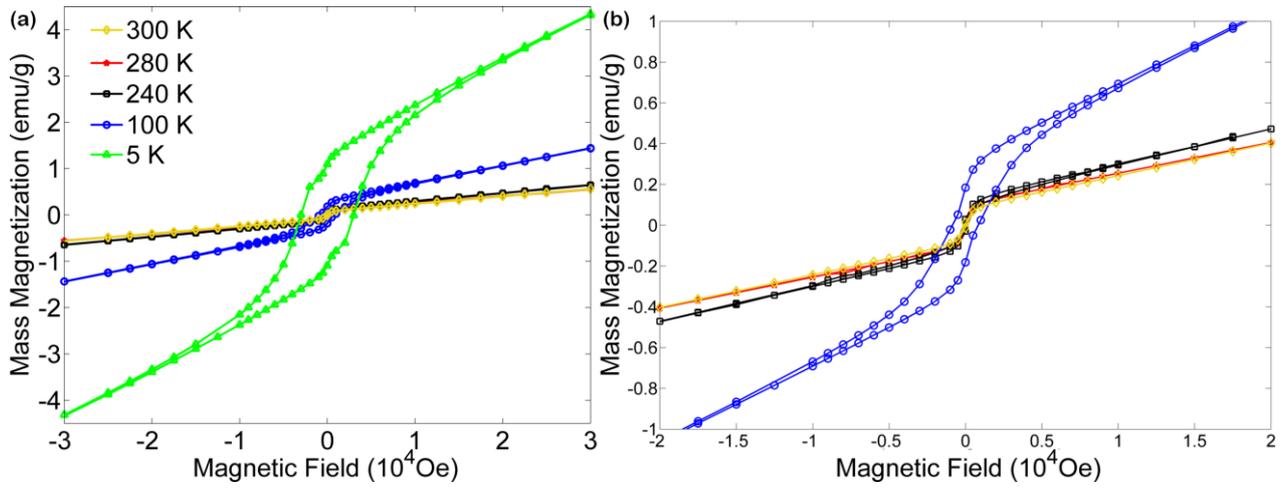

Figure 11

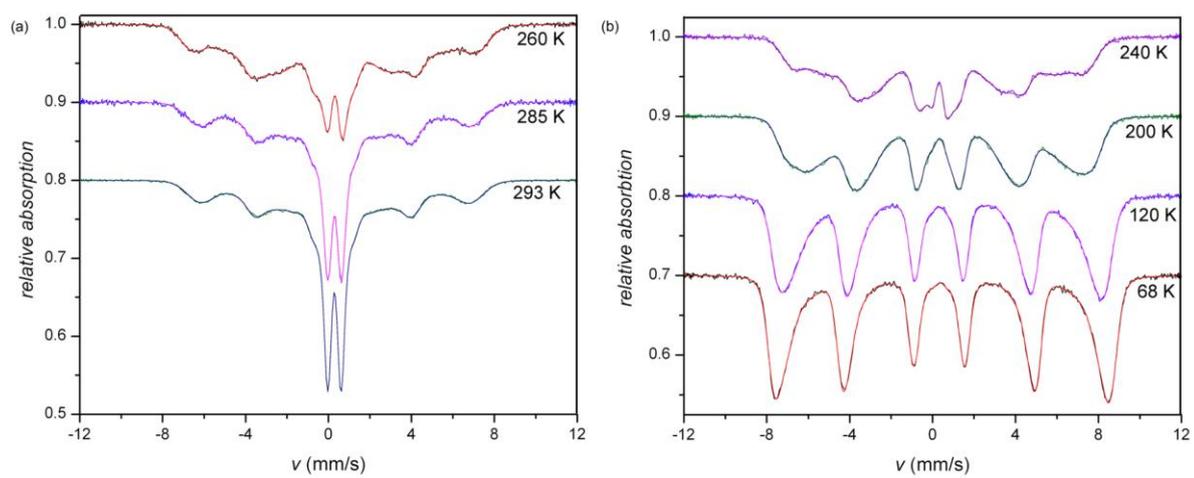

Figure 12

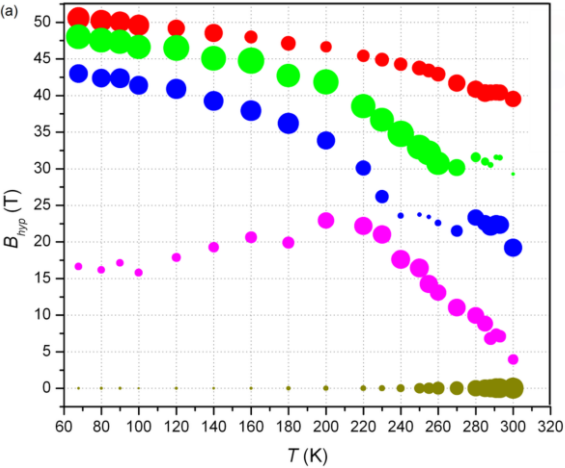 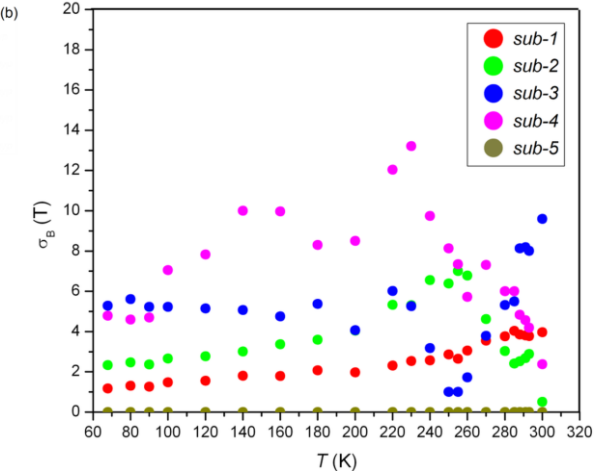

Figure 13

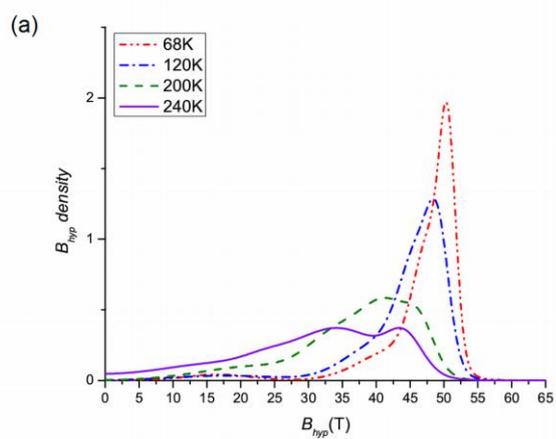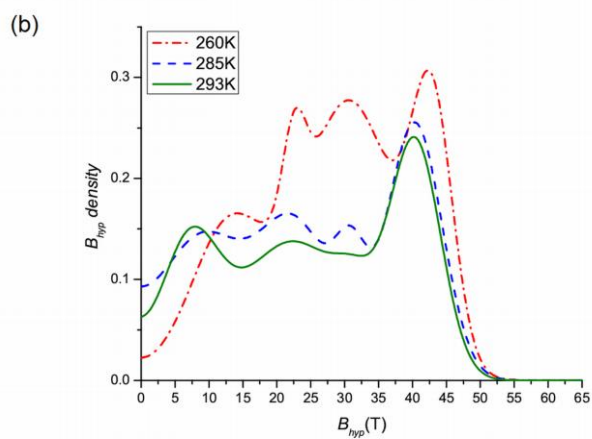